# Fourier Codes and Hartley Codes

H.M. de Oliveira, C.M.F. Barros, R.M. Campello de Souza

*Abstract*—Real-valued block codes are introduced, which are derived from Discrete Fourier Transforms (DFT) and Discrete Hartley Transforms (DHT). These algebraic structures are built from the eigensequences of the transforms. Generator and parity check matrices were computed for codes up to block length *N*=24. They can be viewed as lattices codes so the main parameters (dimension, minimal norm, area of the Voronoi region, density, and centre density) are computed. Particularly, Hamming-Hartley and Golay-Hartley block codes are presented. These codes may possibly help an efficient computation of a DHT/DFT.

## I. INTRODUCTION

Discrete transforms over finite or infinite fields have long been used in the Telecommunication field to achieve different goals. Transforms have significant applications on subjects such as channel coding, cryptography and digital signal and image processing. They were primary conceived to perform an efficient and fast numerical spectral analysis [1], [2]. However, several new applications emerged, which are transformed-based. The better known example is the Discrete Fourier Transform (DFT). Another very rich transform related to the DFT is the Discrete Hartley Transform (DHT) [3], [4]. The DHT has proven over the years to be a powerful tool. A version of the DFT for finite fields was introduced by Pollard in early 70´s [5] and applied as a tool to perform discrete convolutions using integer arithmetic. Some particular transforms, mainly the DHT, have found interesting applications in the optical domain [6], [7]. New digital multiplex systems have been proposed, which are derived from finite field transforms [8]. Recent promising applications of discrete transforms concern the use of transforms to design digital multiplex systems and spread spectrum sequences [9], [10]. This approach exploits orthogonality properties of synchronous multilevel sequences defined over a complex finite field [11] as a tool of spreading sequence design. Additionally, finite field transforms were offered to implement efficient multiple access systems [9], [12].

The uppermost application of discrete transforms is definitely in signal and image compression, where they are the basis of countless algorithms [13]. The decoding of DTMF signals also is always performed on the basis of transforms [14], [15]. Discrete transforms have been successfully applied in error control coding schemes, both in the design of new codes and decoding algorithms [16], [17], [18], [19].

Some cryptographic systems have been devised that exploits discrete transforms [20]. The discrete multitone (DMT) systems are essentially based on DFTs [21]. A related type of modulation, the orthogonal frequency division multiplex (OFDM multicarrier systems), has been effectively applied in digital broadcast and wireless channel communication [22], [23], [24]. They are also a very efficient tool for spectral monitoring, therefore are extensively used in spectral managing [25], [26].

Eigenfunctions of discrete transforms have been one of the focus of several studies [27], [28], [29]. This analysis recently derived new multi-user systems [30]. This paper links the eigenvalues of discrete transforms [31] with the design of block codes defined over the field of real numbers [32], [33]. The structure and parameters of the real codes derived from this approach are presented, and they are put on the lattice codes framework [34], [35], showing that *discrete transforms* have connection, somewhat unexpected, with *lattice theory*.

The codebook of these codes is simply a list of all eigensequences associated with a particular real-valued eigenvalue of the corresponding discrete transform [31]. The generator (*G*) and parity check (*H*) matrices are introduced for such codes, which play a role somewhat similar to the standard *G* and *H* matrices of block codes [36].

The paper is organized as follows. Section II presents the major concepts and makeup of the transformed-based real block codes, with focus on the so-called Fourier codes. Section III shows several codes derived from discrete Hartley transforms (named Hartley codes). Concluding remarks are presented in Section IV.

## II. ARE DISCRETE TRANSFORMS RELATED TO REAL BLOCK CODES?

Let $[W]_N := \left( \exp(-j\frac{2\pi}{N}k.n) \right)$ be the $N \times N$ DFT matrix, where $k,n=0,1,2,\ldots,N-1$. If $x[n]$ is an eigensequence of the linear transform $W$, then its spectrum (Fourier spectrum) is given by:

$$[W]_N x[n] = \lambda x[n]. \quad (1)$$

The four possible eigenvalues of a DFT with blocklength $N$, $\lambda$, are only $\pm\sqrt{N}, \pm j\sqrt{N}$ [29],[30]. Each one of the four eigenvalues engender a linear subspace of eigensequences of blocklength $N$, denoted by $V_N^+, V_N^-, V_N^j, V_N^{-j}$, respectively [30].

The authors are with the Federal University of Pernambuco - UFPE, Signal Processing Group, C.P. 7.800, 50.711-970, Recife - PE, Brazil (e-mail: caio1408@yahoo.com.br, hmo@ufpe.br, ricardo@ufpe.br). This work was partially supported by the Brazilian National Council for Scientific and Technological Development (CNPq) under research grant #301996.

Since we intend to deal solely with "real codes", just the real-valued eigenvalues are of interest. Accordingly, the eigensequences $x[n]$ must hold the relationship

$$[W]_N x[n] = \pm\sqrt{N}.x[n], \quad (2)$$

so that $x[n]([W]_N \mp \sqrt{N}I_N)^T = 0$, where the sign depends on if the eigenvalue is positive or negative, respectively. As a result, the matrix $H^T := [W]_N \mp \sqrt{N}I_N$ plays a role to some extent analogous to the parity check matrix of standard block codes.

$$\text{Here } N-K = rank([W]_N \mp \sqrt{N}I_N). \quad (3)$$

It seems appropriate to adopt a notation corresponding to the standard approach of block codes, that is,

$$H^T = [I_{N-K} \vdots P_\lambda] \quad (4)$$

in such a way that the generator matrix of the real code can be put under the format

$$G = [-P_\lambda^T \vdots I_K]. \quad (5)$$

Two block codes of parameters $[N,K^+]$ and $[N,K^-]$ may be generated, where

$$K^{sgn(\lambda)} = N - rank([W]_N - \lambda I_N), \lambda = \pm\sqrt{N}, \quad (6)$$

where $sgn(.)$ is the usual signal function.

*Example 1.* As a naïve and clarifying example, let us consider the Fourier codes of length $N=4$. We start with the matrix

$$([W]_N - \lambda I_N) = \begin{vmatrix} -1 & 1 & 1 & 1 \\ 1 & -(2+j) & -1 & j \\ 1 & -1 & -1 & -1 \\ 1 & j & -1 & -(2+j) \end{vmatrix} \quad (7)$$

Thus, $K^+ = 2$. In standard echelon form, the "parity check" matrix, is $\begin{vmatrix} 1 & 0 & -1 & -2 \\ 0 & 1 & 0 & -1 \end{vmatrix}$, so we have the following two Fourier codes:

$$F: [4,2] \quad G_4^+ = \begin{vmatrix} 1 & 0 \vdots 1 & 0 \\ 2 & 1 \vdots 0 & 1 \end{vmatrix} \quad (8)$$

$$F: [4,1] \quad G_4^- = |-1 \quad 1 \quad 1 \vdots 1|. \quad (9)$$

Codes conceived by this approach can be interpreted as lattice codes [37], [35].

The corresponding codes engendered by the unitary Hartley transform of length four have the following matrices:

$$H: [4,3] \quad G_4^+ = \begin{vmatrix} 1 & 1 & 0 & 0 \\ 1 & 0 & 1 & 0 \\ 1 & 0 & 0 & 1 \end{vmatrix} \quad (10)$$

$$H: [4,1] \quad G_4^- = |-1 \quad 1 \quad 1 \vdots 1|. \quad (11)$$

Incidentally, these generator matrices corresponds exactly to the lattices $D_3$ (checkerboard lattice) and $A_1$, the best packing in dimension 3 and 1, respectively [34], [35]. It seems to be appealing to make use of the unitary form of the DFT/DHT operator. However, lattice codes thus derived are tantamount to those obtained from the standard DHT. Fourier codes parameters were computed for blocklengths up to 24 and the results are summarized in Table I. The density achieved by the corresponding sphere packing is indicated.

*TABLE I. Fourier lattice Codes on an N-dimensional Euclidean space. K is the dimension, μ the minimal norm, and Δ the density of the associated lattice packing.*

| N | $K^+$ | $\mu^+$ | $\Delta^+$ | $K^-$ | $\mu^-$ | $\Delta^-$ |
|---|---|---|---|---|---|---|
| 3 | 1 | 9.4641 | 1 | 1 | 2.5359 | 1 |
| 4 | 2 | 2 | 0.55536 | 1 | 4 | 1 |
| 5 | 2 | 4 | 0.82582 | 1 | 5.5279 | 1 |
| 6 | 2 | 3.5505 | 0.56921 | 2 | 3.5505 | 0.56921 |
| 7 | 2 | 2.8931 | 0.37722 | 2 | 2.4577 | 0.44763 |
| 8 | 3 | 3 | 0.26029 | 2 | 2.3431 | 0.42505 |
| 9 | 3 | 2.7852 | 0.26073 | 2 | 2.5014 | 0.43446 |
| 10 | 3 | 2.7906 | 0.25213 | 3 | 6.8377 | 0.40055 |
| 11 | 3 | 2.9203 | 0.2483 | 3 | 5.4021 | 0.42541 |
| 12 | 4 | 4 | 0.11577 | 3 | 4.906 | 0.44661 |
| 13 | 4 | 3.4628 | 0.12822 | 3 | 4.2503 | 0.39718 |
| 14 | 4 | 3.5376 | 0.15731 | 3 | 3.8717 | 0.30403 |
| 15 | 4 | 3.4413 | 0.15837 | 4 | 7.418 | 0.21238 |
| 16 | 4 | 4.2625 | 0.1412 | 3 | 3.2543 | 0.27218 |
| 17 | 5 | 5.3824 | 0.089659 | 4 | 2.9451 | 0.063149 |
| 18 | 5 | 4.8172 | 0.096808 | 4 | 3.0745 | 0.069407 |
| 19 | 5 | 4.7151 | 0.11403 | 4 | 3.304 | 0.088451 |
| 20 | 5 | 5.0325 | 0.13599 | 4 | 4.1055 | 0.15164 |
| 21 | 5 | 4.9701 | 0.10052 | 4 | 3.6455 | 0.14034 |
| 22 | 6 | 7.2991 | 0.071307 | 5 | 8.387 | 0.10898 |
| 23 | 6 | 5.0715 | 0.034921 | 5 | 8.6131 | 0.15042 |
| 24 | 6 | 4.0909 | 0.021066 | 5 | 5.777 | 0.1134 |

Further parameters for the Fourier Codes defined over the Euclidean space are shown in Table II, including the determinant of the Gram matrix [35], and their centre densities. In the next section, codes derived from the discrete Hartley transform (DHT) are introduced. Long codes can easily be designed by using long blocklength discrete transforms.

*TABLE II. Fourier Codes on an N-dimensional Euclidean space. Further parameters: δ is the centre density of the associated lattice packing, and det(Λ) is the volume of the Voronoi region.*

| N | $K^+$ | $\delta^+$ | det $\Lambda^+$ | $K^-$ | $\delta^-$ | det $\Lambda^-$ |
|---|---|---|---|---|---|---|
| 3 | 1 | 0.5 | 3.0764 | 1 | 0.5 | 1.5925 |
| 4 | 2 | 0.17678 | 2.8284 | 1 | 0.5 | 2 |
| 5 | 2 | 0.26287 | 3.8042 | 1 | 0.5 | 2.3511 |
| 6 | 2 | 0.18119 | 4.899 | 2 | 0.18119 | 4.899 |
| 7 | 2 | 0.12007 | 6.0237 | 2 | 0.14249 | 4.3122 |
| 8 | 3 | 0.06214 | 10.453 | 2 | 0.1353 | 4.3296 |
| 9 | 3 | 0.062245 | 9.3343 | 2 | 0.13829 | 4.5221 |
| 10 | 3 | 0.060192 | 9.6812 | 3 | 0.095625 | 23.373 |
| 11 | 3 | 0.059278 | 10.524 | 3 | 0.10156 | 15.454 |
| 12 | 4 | 0.023459 | 42.628 | 3 | 0.10662 | 12.74 |
| 13 | 4 | 0.025982 | 28.844 | 3 | 0.094819 | 11.552 |
| 14 | 4 | 0.031878 | 24.536 | 3 | 0.072582 | 13.12 |
| 15 | 4 | 0.032093 | 23.062 | 4 | 0.043038 | 79.91 |
| 16 | 4 | 0.028614 | 39.685 | 3 | 0.064979 | 11.294 |
| 17 | 5 | 0.017033 | 123.31 | 4 | 0.012797 | 42.364 |
| 18 | 5 | 0.018391 | 86.543 | 4 | 0.014065 | 42.005 |
| 19 | 5 | 0.021663 | 69.642 | 4 | 0.017924 | 38.066 |
| 20 | 5 | 0.025835 | 68.723 | 4 | 0.030728 | 34.283 |
| 21 | 5 | 0.019096 | 90.122 | 4 | 0.028438 | 29.207 |
| 22 | 6 | 0.013799 | 440.35 | 5 | 0.020704 | 307.47 |
| 23 | 6 | 0.0067576 | 301.61 | 5 | 0.028577 | 238.09 |
| 24 | 6 | 0.0040764 | 262.42 | 5 | 0.021544 | 116.35 |

## III. BLOCK-CODES LIKE STRUCTURES DERIVED FROM HARTLEY TRANSFORMS: A TABLE OF HARTLEY CODES

Let $[H]_N := \left( cas(\frac{2\pi}{N} k.n) \right)$ $n,k=0,1,\ldots,N-1$ be the $N \times N$ DHT matrix. The (real) casoidal Hartley kernel is $cas(x):=cos(x)+sin(x)$ as usual. Then the eigensequences $x[n]$ of the DHT must hold $[H]_N x[n] = \pm \sqrt{N} . x[n]$, so that $x[n]([H]_N \mp \sqrt{N} I_N)^T = 0$, where the sign depends on if the eigenvalue is positive or negative, respectively. Again, the matrix $H^T := [H]_N \mp \sqrt{N} I_N$ plays a role somewhat similar to the parity check matrix of eigensequences. Here again

$$N-K=rank([W]_N \mp \sqrt{N} I_N). \quad (12)$$

Hartley codes have dimensions given by

$$K^+ = \left\lceil \frac{N}{2} \right\rceil + \delta_{N \bmod 4,0} \text{ and } K^- = \left\lfloor \frac{N}{2} \right\rfloor - \delta_{N \bmod 4,0}, \quad (13)$$

where $\delta_{k,l}$ denotes the Kronecker symbol. It is worthwhile to remark that the "rate" $K^\pm/N$ of these codes is always about a half.

It is worthwhile to mention that the lattices $\Lambda$ and $\Lambda^\perp$ corresponding to the codes $[N,K^+]$ and $[N,K^-]$ are dual lattices. Furthermore, the computational complexity required to compute the parameters of Hartley codes is much higher than that required by Fourier codes of the same length.

*Corollary*. The matrix of the DHT of blocklength $N$ has only eigenvalues $+1$ and $-1$ and they are repeated $\left\lceil \frac{N}{2} \right\rceil + \delta_{N \bmod 4,0}$ times and $\left\lfloor \frac{N}{2} \right\rfloor - \delta_{N \bmod 4,0}$ times, respectively.

The generator and the parity check matrices of the real block code of length $N$ associated with the eigenvalue $\lambda$ of the DFT (DHT) will be denoted by $_s G_N^{sgn(\lambda)}$ and $_s H_N^{sgn(\lambda)}$, where $s$ denote Fourier or Hartley code $s \in \{F, H\}$, and $sgn(\lambda) \in \{+, -\}$.

*Example 2*. As a naïve and illustrative example, let us consider the Hartley codes of block length 4.

$$_H G_4^- = \begin{vmatrix} 1 & 0 & 0 & 1 \\ 0 & 1 & 0 & -1 \\ 0 & 0 & 1 & -1 \end{vmatrix} \text{ and } (_H H_4^-)^T = \begin{vmatrix} -1 & 1 & 1 & 1 \end{vmatrix}.$$

Let also build a Hartley code of block length $N=7$ associated with the eigenvalue $-\sqrt{7}$. The generator matrix is:

$$_H G_7^- = \begin{bmatrix} 1 & 0 & 0 & 0 & 0.2540 & 0.4240 & 0.9678 \\ 0 & 1 & 0 & 0 & -0.6697 & -0.2831 & -0.0472 \\ 0 & 0 & 1 & 0 & 1.2068 & -0.4899 & -1.7169 \\ 0 & 0 & 0 & 1 & -0.4629 & 0.2270 & -0.7641 \end{bmatrix}$$

A particular and remarkable case occurs for blocklengths for which there exists eigensequences with integer components. We have shown that this happens when $N=m^2$ and this is illustrated in the example in the sequel.

*Example 3*. For instance, take the case of the DHT $N=9$ and select the eigenvalue $\lambda=1$. The Hartley code $[9,5]$ has generator matrix $_H G_9^+$ given by

| 6.2150 | 6.1346 | 2.7832 | 2.5123 | 1 | 0 | 0 | 0 | 0 |
|---|---|---|---|---|---|---|---|---|
| -2.0501 | -2.7832 | -0.3139 | -2.0030 | 0 | 1 | 0 | 0 | 0 |
| 1.0000 | 0.0000 | 0.0000 | 1.0000 | 0 | 0 | 1 | 0 | 0 |
| -3.8490 | -5.1346 | -2.7832 | -0.7802 | 0 | 0 | 0 | 1 | 0 |
| 2.6840 | 2.7832 | 1.3139 | 0.2710 | 0 | 0 | 0 | 0 | 1 |

so there are sequences with all-integer components such as:

$[1\ 0\ 0\ 1\ 0\ 0\ 1\ 0\ 0]$, $[10\ 1\ 1\ 7\ 1\ 1\ 7\ 1\ 1]$, $[10\ 1\ 1\ 1\ 1\ 1\ 1\ 1\ 1]$.

*Example 4*. The Hamming-Hartley and Golay-Hartley codes. The generator matrix of the real Hamming-Hartley code $[N,K^+]=[7,4]$ is

$$_H G_7^+ = \begin{vmatrix} 3.9372 & 3.81064 & 1.66971 & 1 & 0 & 0 & 0 \\ 1.82300 & 1 & 1 & 0 & 1 & 0 & 0 \\ -4.75175 & -6.31546 & -2.50481 & 0 & 0 & 1 & 0 \\ 2.63705 & 2.50481 & 0.83511 & 0 & 0 & 0 & 1 \end{vmatrix}$$

The generator matrix of the real Golay-Hartley code $[N,K^+]=[23,12]$ is given in Appendix.

TABLE III. Hartley Codes on an $N$-dimensional Euclidean space. $K$ is the dimension, $\mu$ the minimal norm, and $\Delta$ the density of the associated lattice packing.

| N | K⁺ | μ⁺ | Δ⁺ | K⁻ | μ⁻ | Δ⁻ |
|---|---|---|---|---|---|---|
| 3 | 2 | 2 | 0.7221 | 1 | 2.5359 | 1 |
| 4 | 3 | 2 | 0.74048 | 1 | 4 | 1 |
| 5 | 3 | 4 | 0.42552 | 2 | 1.5814 | 0.4614 |
| 6 | 3 | 4.9148 | 0.21497 | 3 | 1.3924 | 0.24623 |
| 7 | 4 | 3.3937 | 0.19977 | 3 | 1.5515 | 0.22407 |
| 8 | 5 | 2 | 0.062949 | 3 | 2.3431 | 0.30672 |
| 9 | 5 | 3 | 0.054289 | 4 | 1.7306 | 0.12155 |
| 10 | 5 | 5.2063 | 0.075916 | 5 | 1.8957 | 0.077169 |
| 11 | 6 | 3.7751 | 0.052394 | 5 | 2.6115 | 0.14192 |
| 12 | 7 | 2.5744 | 0.014886 | 5 | 2.2205 | 0.066449 |
| 13 | 7 | 3.0709 | 0.0099451 | 6 | 1.8702 | 0.024582 |
| 14 | 6 | 2 | 0.010263 | 7 | 2.2839 | 0.02135 |
| 15 | 8 | 4.1987 | 0.015084 | 7 | 2.1276 | 0.014352 |
| 16 | 9 | 2.7463 | 0.0022665 | 7 | 2.2606 | 0.012029 |
| 17 | 9 | 3.3358 | 0.0021189 | 8 | 1.9687 | 0.00391907 |
| 18 | 9 | 4.7288 | 0.0033659 | 9 | 2.5477 | 0.0047677 |
| 23 | 12 | - | - | 11 | - | - |

A glimpse on the properties of these lattices may be a bit disappointing, because they do not provide dense packings. Nonetheless, it is significant to stress at this point, that the aim of this paper is not to find dense packings or thin coverages. After all, this task is recognized as an exceptionally hard one.

*TABLE IV. Hartley Codes on an N-dimensional Euclidean space. Further parameters: δ is the centre density of the associated lattice packing, and det(Λ) is the volume of the Voronoi region.*

| N  | $K^+$ | $\delta^+$ | det $\Lambda^+$ | $K^-$ | $\delta^-$ | det $\Lambda^-$ |
|----|-------|-----------|-----------------|-------|------------|-----------------|
| 3  | 2     | 0.22985   | 2.1753          | 1     | 0.5        | 1.5925          |
| 4  | 3     | 0.17678   | 2               | 1     | 0.5        | 2               |
| 5  | 3     | 0.10159   | 9.8438          | 2     | 0.14687    | 2.6919          |
| 6  | 3     | 0.051319  | 26.54           | 3     | 0.058782   | 3.494           |
| 7  | 4     | 0.040483  | 17.781          | 3     | 0.053494   | 4.5157          |
| 8  | 5     | 0.011959  | 14.782          | 3     | 0.073223   | 6.1229          |
| 9  | 5     | 0.010314  | 47.233          | 4     | 0.024631   | 7.5998          |
| 10 | 5     | 0.014422  | 134.01          | 5     | 0.01466    | 10.547          |
| 11 | 6     | 0.010139  | 82.915          | 5     | 0.026962   | 12.774          |
| 12 | 7     | 0.0031505 | 67.882          | 5     | 0.012624   | 18.189          |
| 13 | 7     | 0.0021049 | 188.36          | 6     | 0.0047569  | 21.488          |
| 14 | 6     | 0.001986  | 62.94           | 7     | 0.0045188  | 31.127          |
| 15 | 8     | 0.003716  | 326.64          | 7     | 0.030375   | 36.13           |
| 16 | 9     | 0.00068728| 267.95          | 7     | 0.0025442  | 53.298          |
| 17 | 9     | 0.00064239| 687.59          | 8     | 0.00006449 | 60.769          |
| 18 | 9     | 0.00000161| 2081.2          | 9     | 0.00005246 | 90.866          |
| 23 | 12    | -         | -               | 11    | -          | -               |

## IV. CONCLUSIONS

The aim of this paper is to launch an alternative approach for the designing of block codes over the field of real numbers. The most relevant issue here is to establish a structure in terms of eigensequences, which may possibly help an efficient computation of a DHT/DFT by partitioning the transform into "sub-transforms" defined over the invariant spaces $V_N^+, V_N^-$. Long codes can easily be designed by using long blocklength discrete transforms. In spite of the fact that this paper is only a preliminary investigation on the link between discrete transforms and lattices, it opens a path for unusual applications of the lattice theory. In particular, Discrete Hartley transform matrices seem to have a strong connection with lattice construction. The structure of these codes was briefly examined, but their performance over noisy channels was not addressed [38]. For instance, for the additive white Gaussian channel not only the lattice density plays a role on the error performance, but also the number of neighbours (contact number). At any rate, it must be said that these are not codes intended for the Gaussian additive noise channel. Further transforms close related to the DHT such as the Generalized-DHT should also be scrutinized to find further lattice codes. The particular mathematical arrangement behind such codes may potentially assist the implementation of fast transforms using trellis, a topic which is currently under investigation. Finally, the search for finite groups that lie hidden in these lattices is also another subject for upcoming researches.

ACKNOWLEDGEMENTS. The authors are indebted to Mr. Gilson Jerônimo da Silva Jr. for valuable discussions.

## REFERENCES

[1] J.W. Cooley. How the FFT gained acceptance, *IEEE Signal Process. Mag.*, Jan., pp.10-13, 1992.

[2] H.M. de Oliveira, R.M. Campello Souza. A Fast Algorithm for Computing the Hartley/Fourier Spectrum *Anais da Academia Brasileira de Ciências.* Rio de Jan., vol. 73, pp.468-468. 2001.

[3] R.N. Bracewell. The Discrete Hartley Transform, *J. Opt. Soc. Amer.*, vol. 73, pp. 1832-1835, 1983.

[4] R.N. Bracewell. *The Hartley Transform*. Oxford University Press, 1986.

[5] J. M. Pollard. The Fast Fourier Transform in a Finite Field, *Math. Comput.*, vol.25, n.114, pp. 365-374, Apr. 1971.

[6] R.N. Bracewell. Aspects of the Hartley Transform. *Proc. of the IEEE* vol. 82. pp. 381-387, Mar., 1994.

[7] K.J. Olejniczac, G.T. Heydt. Scanning the Special Section on the Hartley Transform, *Proc. of the IEEE*, vol.82, pp.372-380, Mar, 1994.

[8] H.M. de Oliveira, R.M. Campello de Souza, A.N. Kauffman. Efficient Multiplex for Band-Limited Channels: Galois-Field Multiple Access. WCC. INRIA. Paris. *Proc. of the Workshop on Coding and Cryptography'99*, 1999, pp. 235-241.

[9] H.M. de Oliveira, R.M. Campello de Souza, A.N. Kauffman. Orthogonal Multilevel Spreading Sequence Design. In: 5th *International Symposium on Comm. theory and Applications*, ISCTA, 1999, Ambleside, 1999, vol.1, pp.206-208.

[10] H.M. de Oliveira. J.P.C.L. Miranda. R.M. Campello de Souza. Spread-Spectrum Based on Finite Field Fourier Transforms. Proc. of the ICSECIT (*Int. Conf. on System Engineering. Comm. and. Info. Technol.*). Punta Arenas, 2001.

[11] R.M. Campello de Souza, H.M. de Oliveira, A.N. Kauffman. The Hartley Transform over a Finite Field. *Revista da Sociedade Brasileira de Telecomunicações*. vol. 14., n.1., pp. 46-54, 1999.

[12] J.P.C.L. Miranda, H.M. de Oliveira. On Galois-Division Multiple Access Systems: Figures of Merit and Performance Evaluation. *XIX Simpósio Brasileiro de Telecomunicações*, 2001, Fortaleza CE.

[13] J. Kovacevic, M. Vetterli. Transform Coding: Past, Present and Future, *IEEE Signal Process. Magazine*, vol.18, n.5, Sept. 2001.

[14] J.B. Lima, R.M. Campello de Souza, H.M. de Oliveira, M.M.C. Souza. Decodificação de Sinais DTMF Via Transformada Aritmética de Fourier. *XXI Simpósio Brasileiro de Telecomunicações*, SBrT'04. Belém, Pará, 6-9 Sept., 2004.

[15] J.B. Lima, R.M. Campello de Souza, H.M. de Oliveira, M.M.C. Souza. Faster DTMF Decoding. *Lecture Notes in Computer Science*. LNCS 3124, J.N. Souza, P. Dini, P. Lorenz eds. Heidelberg: Springer Verlag. 2004, vol. 1, pp.510-515, 2004.

[16] R.E. Blahut. Transform Techniques for Error-Control Codes. *IBM J. Res. Dev.* vol. 23. pp. 299-315. May 1979.

[17] J.-L. Wu,. J. Shiu. Discrete Hartley Transform in Error Control Coding. *IEEE Trans. Acoust. Speech. Signal Processing*, vol. ASSP-39, pp. 2356-2359, Oct. 1991.

[18] F. Marvasti, M. Nafie. Using FFT for Error Correction Decoders, *IEE Colloquium on DSP Applications in Communication Systems*, pp. 9/1-9/4, London, UK Mar 1993.

[19] J.-L. Wu, J. Shin. Discrete Cosine Transform in Error Control Coding. *IEEE Trans. on Communications*, vol. 43, pp. 1857-1861 May 1995.

[20] J.L. Massey. The Discrete Fourier Transform in Coding and Cryptography, *IEEE Inform. Theory Workshop*, ITW 98, San Diego, 1998.

[21] S. Haykin. *Communication Systems*, 4/e, Wiley, 2001.

[22] S.B. Weinstein, P.M. Ebert. Data Transmission by Frequency Division Multiplexing Using the Discrete Fourier Transform. *IEEE Transactions on Communication Technology*, vol. 19, n.5, Oct. 1971.

[23] A. Bohdanowicz, C. Bos, M. Ditzel, W. Serdijin, G. Janseen, Deprettere. Ed. *Wireless Link using OFDM Modulation: Performance Prediction, Modeling and Implementation.* Delft University of Technology, 1998.

[24] L.J. Cimini Jr, Y. Li. Orthogonal Frequency Division Multiplexing for Wireless Communications, tutorial notes TU18, *Int. Conf. on Comm.*, British Columbia, Canada, June, 1999.

[25] K. Song, S. Chung, G. Ginis, J. Cioffi. Dynamic Spectrum Management for Next-generation DSL Systems, *IEEE Comm. Mag.*, vol.40, pp.101-109, 2002.

[26] R. Cendrillon, M. Moonen. Iterative Spectrum Management for Digital Subscriber Lines, *IEEE Int. Conf. on Comm.*, ICC, 2005, pp.1937-1941.

[27] S-C. Pei, J.-J. Ding. Eigenfunctions of Linear Canonical Transforms, *IEEE Trans. on Signal Proc.*, vol.50, n.1, Jan.,pp.11-26, 2002.

[28] C-C. Tseng, Eigenvalues and Eigenvectors of Generalized DFT, Generalized DHT, DCT-IV and DST-IV Matrices, IEEE Trans. on Signal Process., vol.50, April, pp.866-877, 2002.

[29] L.R. Soares, H.M. de Oliveira, R.J.S. Cintra, R.M.C. Souza. Fourier Eigenfunctions, Uncertainty Gabor Principle and Isoresolution Wavelets, *XX Simpósio Bras. de Telecomunicações*, Rio de Janeiro, 5-8 Oct., 2003. Available: http://www2.ee.ufpe.br/codec/isoresolution_vf.pdf

[30] R.M. Campello de Souza, H.M. de Oliveira. Eigensequences for Multiuser Communication over the Real Adder Channel. IEEE *VI International Telecommunications Symposium* (ITS2006). Sept. 3-6. Fortaleza. Brazil. Available: http://www2.ee.ufpe.br/codec/ITS_SIDFT.pdf

[31] J.B. Lima, R.M. Campello Souza, Uma Técnica de Múltiplo Acesso Baseada na Auto-estrutura das Transformadas Trigonométricas, submetido ao *XXV SBrT*, 2007.

[32] T. Marshall Jr. Coding of Real-Number Sequences for Error Correction: A Digital Signal Processing Problem, *IEEE J. on Selected Areas in Communications*, vol. 2, pp.381-392, Mar., 1984.

[33] L.F. Borodin. Correcting Codes over the Real Number Field. *J. Commun. Technol.* vol. 48, n.8, pp. 896-902, 2003.

[34] T.M. Thompson. *From Error-Correcting Codes Through Sphere Packings to Simple Groups*. The Math. Assoc. of America, 1983.

[35] J.H. Conway. N.J.A. Sloane. *Sphere Packings. Lattices and Groups*. NY: Springer-Verlag, 1988.

[36] S. Wicker. *Error Control Systems for Digital Communication and Storage*, Prentice-Hall, 1995.

[37] N.J.A. Sloane, *The Sphere Packing Problem*, 1998 Shannon Lecture, 1998.

[38] H.M. de Oliveira, G. Battail. Performance of Lattice Codes over the Gaussian Channel. *Annales des Télécommunications*. vol. 47, n.7-8, pp.293-305, 1992.

**APPENDIX** Real parity check matrix of the Golay-Hartley code, $_H H^+_{23}$.

| | | | | | | | | | | | | | | | | | | | | | | |
|---|---|---|---|---|---|---|---|---|---|---|---|---|---|---|---|---|---|---|---|---|---|---|
| 1 | 0 | 0 | 0 | 0 | 0 | 0 | 0 | 0 | 0 | 0 | 0 | 0.0707 | 0.1328 | 0.1806 | 0.2052 | 0.2576 | 0.2725 | 0.3446 | 0.3546 | 0.4865 | 0.4931 | 0.9975 |
| 0 | 1 | 0 | 0 | 0 | 0 | 0 | 0 | 0 | 0 | 0 | 0 | -0.2096 | -0.3738 | -0.4600 | -0.4311 | -0.4377 | -0.2763 | -0.2355 | 0.0703 | 0.0802 | 0.8307 | 0.4427 |
| 0 | 0 | 1 | 0 | 0 | 0 | 0 | 0 | 0 | 0 | 0 | 0 | 0.3139 | 0.4952 | 0.4639 | 0.1856 | -0.0295 | -0.4240 | -0.4346 | -0.6661 | 0.0749 | -0.3157 | -0.6636 |
| 0 | 0 | 0 | 1 | 0 | 0 | 0 | 0 | 0 | 0 | 0 | 0 | -0.3631 | -0.4411 | -0.1470 | 0.3930 | 0.5700 | 0.6670 | 0.0734 | 0.2235 | -0.6462 | -1.1716 | -0.1579 |
| 0 | 0 | 0 | 0 | 1 | 0 | 0 | 0 | 0 | 0 | 0 | 0 | 0.4208 | 0.3418 | -0.1762 | -0.6303 | -0.1794 | 0.2343 | 1.1547 | 0.1737 | -1.0834 | -0.7214 | -0.5346 |
| 0 | 0 | 0 | 0 | 0 | 1 | 0 | 0 | 0 | 0 | 0 | 0 | -0.4200 | -0.0859 | 0.5313 | 0.5639 | -0.3692 | 0.0609 | 0.1053 | -0.1063 | 0.0355 | -0.8583 | -0.4572 |
| 0 | 0 | 0 | 0 | 0 | 0 | 1 | 0 | 0 | 0 | 0 | 0 | 0.4704 | -0.0758 | -0.4440 | 0.1100 | 1.1463 | -0.2617 | -1.2212 | 0.2180 | 0.2166 | -0.4859 | -0.6726 |
| 0 | 0 | 0 | 0 | 0 | 0 | 0 | 1 | 0 | 0 | 0 | 0 | -0.4108 | 0.4409 | 0.4098 | 0.0483 | -0.0264 | -0.1150 | -0.3806 | -0.5899 | 0.5268 | -0.2818 | -0.6212 |
| 0 | 0 | 0 | 0 | 0 | 0 | 0 | 0 | 1 | 0 | 0 | 0 | 0.4964 | -0.4527 | 0.5468 | 0.4790 | -1.2225 | 0.1046 | 0.1456 | -0.6493 | 0.1306 | 0.1316 | -0.7100 |
| 0 | 0 | 0 | 0 | 0 | 0 | 0 | 0 | 0 | 1 | 0 | 0 | -0.3338 | 1.1216 | -0.3070 | -0.7009 | 0.4065 | -0.8335 | 0.5190 | -0.3977 | -0.1826 | 0.3034 | -0.5947 |
| 0 | 0 | 0 | 0 | 0 | 0 | 0 | 0 | 0 | 0 | 1 | 0 | 0.7645 | -0.5050 | -0.1304 | -0.4541 | 0.0619 | -0.0811 | -0.2977 | 0.3274 | -0.5944 | 0.4580 | -0.5491 |
| 0 | 0 | 0 | 0 | 0 | 0 | 0 | 0 | 0 | 0 | 0 | 1 | -0.1386 | -0.2349 | -0.3342 | 0.2470 | -0.4127 | 0.3455 | -0.4255 | 0.3412 | -0.3778 | 0.2531 | -0.2632 |